\documentclass[preprint,prl,nobibnotes,altaffilletter,amsmath,amssymb,amsfonts]{revtex4}

\usepackage{inputenc}
\inputencoding{ansinew}
\usepackage[T1]{fontenc}
\usepackage{graphicx}
\usepackage{palatino}
\usepackage{mathpazo}
\usepackage{bm}
\usepackage{color}

\begin{document}

\title{The consequence of excess configurational entropy on fragility: the case of a polymer/oligomer blend}

\author{C. Dalle-Ferrier$^{1}$}
\author{S. Simon $^{1,2}$}
\author{W. Zheng$^{2}$}
\author{P. Badrinarayanan$^{2}$}
\author{T. Fennell$^{3}$}
\author{B. Frick$^{3}$}
\author{J.M. Zanotti$^{4}$}
\author{C. Alba-Simionesco$^{1,4}$}
\affiliation{$^{1}$Laboratoire de Chimie Physique, UMR 8000,
B{\^a}timent
349, Universit\'e, Paris-Sud, 91405 Orsay, France \\
$^{2}$ Department of Chemical Engineering, Texas Tech University,
Lubbock, TX 79409, USA\\
$^{3}$ Institut Laue-Langevin, 6 rue Jules Horowitz, 38190
Grenoble, France\\
$^{4}$ Laboratoire Léon Brillouin, UMR 12, CEA-CNRS, 91191- Saclay,
France}

\begin{abstract}
By taking advantage of the molecular weight dependence of the glass
transition of polymers and their ability to form perfectly miscible
blends, we propose a way to modify the fragility of a system, from
fragile to strong, keeping the same glass properties, \emph{i.e.}
vibrational density of states, mean-square displacement and local
structure. Both slow and fast dynamics are investigated by
calorimetry and neutron scattering in an athermal
polystyrene/oligomer blend, and compared to those of a pure 17-mer
polystyrene considered to be a reference, of same $T_g$ . Whereas
the blend and the pure 17-mer have the same heat capacity in the
glass and in the liquid, their fragilities differ strongly. This
difference in fragility is related to an extra configurational
entropy created by the mixing process and acting at a scale much
larger than the interchain distance, without affecting the fast
dynamics and the structure of the glass.
\end{abstract}

\maketitle{}

The origin of the dramatic slowing down of dynamics in molecular
liquid and polymer on the approach to the glass transition is still
a hotly debated question in condensed matter physics. Viscosity and
relaxation time show an increase of more than ten orders of
magnitude while temperature decreases by a few tens of degrees. It
comes with a rapid decrease of the configurational entropy defined
as the excess entropy of the metastable liquid over the crystal (as
suggested by Adam and Gibbs \cite{Adam65}), supporting the
hypothesis of an underlying thermodynamic transition. The relaxation
time $\tau_\alpha$ shows different temperature dependences in
different systems, that can be quantified at the glass transition
temperature $T_g$ via the isobaric fragility, introduced by Angell
\cite{Angell85} $m_P=\left(\frac{\partial
\log_{10}(\tau_\alpha/\tau_0)}{\partial T_g/T}\right)_{T=T_g}$, with
$\tau_0$=1s. The fragility values range from 20 for so-called strong
systems such as silica to more than 200 for some fragile polymeric
systems. The classification of systems according to this index arose
from an attempt to understand the universal slowing down and the
glass formation itself. It is however difficult to extract a generic
description by comparing chemically very different systems that as a
result show quite specific relaxation behavior. Various methods have
been proposed in the literature to maintain constant intermolecular
interactions while changing other properties such as fragility. For
example, a single molecular liquid or polymer may be studied at
different pressures, thereby changing the glass transition in the
same chemical system \cite{Frick03}. Another very effective way to
tune fragility without changing the chemistry is to vary the chain
length of a polymer. This affects $T_g$, $m_P$, heat capacity jump,
density, mean-square displacement (MSD), Boson peak (BP) and other
properties related to the glass transition \cite{Frick03,Ding04}. We
propose here a third approach, changing polydispersity $I_p$. We mix
a high molecular weight $M_n$ polystyrene (PS) with its oligomer
creating an athermal polymer/oligomer blend and compare its behavior
with a monodisperse reference sample of exactly same $T_g$ as the
blend. These two samples are very similar at the local scale, as
will be shown from structural measurements in this Letter. Having
such similar samples can only be achieved using polymers of the same
chemical nature. Thanks to the effect of $M_n$ on their physical
properties, the reference of a pure system such as the 17-mer is
available while it is not in the usual polymer-polymer or molecular
liquids mixtures that have been extensively studied in the
literature
\cite{Tyagi06,Smith06,Robertson04,Duvvuri04,Bohmer93,Zhao08} for
other purposes. A comparative study of these two samples is an ideal
way to better understand fragility while minimally affecting the
system. We present here an extensive study of their thermodynamics,
structure and slow and fast dynamics.

Differential scanning calorimetry and elastic and inelastic neutron
scattering are used to measure the heat capacity, the structure
factor $S(Q)$, the MSD at the nanosecond timescale and the BP in the
blend and in the 17-mer. We have focused on the structure and
dynamics at the local scale corresponding to the usual wave vectors
range studied for glass transition, from 0.5 to 2~\AA$^{-1}$. We
have found that neither thermodynamics nor local structure nor fast
dynamics and vibrations differ in the two samples, although the
fragility is quite different. We propose to rationalize this
difference in fragility via an excess configurational entropy of the
polymer/oligomer blend compared to the monodisperse sample, and to
establish a direct link between these two quantities at a
lengthscale much larger than the interchain distance.

Narrow $M_n$ distribution polystyrene samples (hydrogenated, labeled
$h$, and deuterated, $d$) were purchased from Polymer Standards
Service. All sample characteristics are summarized in table
\ref{TableSample}. A bimodal polystyrene blend has been prepared
with 62$\%$ by weight of PSh92k and 38$\%$ of PSh750) thus obtaining
the same $T_g$ as for the PSh1790. A fully deuterated blend (62$\%$
PSd110k and 38$\%$ PSd800) has also been prepared as well as the
corresponding pure sample PSd2300. Calorimetry experiments have been
performed on the blend and on its pure components, leading to the
conclusion that the PS/oligomer blend is athermal, \emph{i.e.} shows
no enthalpy of mixing, consistent with results on similar systems
\cite{Zheng08,Huang03}. The measurements were performed on a Mettler
Toledo DSC by cooling from above the nominal $T_g$ at different
rates ranging from 0.03 to 30 K.min$^{-1}$ followed by a heating
scan at 10 K.min$^{-1}$. The limiting fictive temperature $T_f'$,
which is equivalent to $T_g$ and defined as the intercept of the
extrapolated liquid and glass lines obtained on heating
\cite{McKenna02}, was calculated using the instrumental software.
All of the neutron scattering experiments were realized using hollow
cylindrical aluminium cells containing a film of the sample wrapped
in an aluminium foil. The films were prepared under vacuum with a
thickness corresponding to a 10\% scattering of the incident beam.
Inelastic neutron scattering measurements were performed on PSh1790
and the fully hydrogenated blend using the time-of-flight instrument
Mib\'{e}mol at the Laboratoire Leon Brillouin (Saclay, France), at
an incoming wavelength of $6$~\AA~and a Full Width Half Maximum
(FWHM) resolution around $100~\mu eV$. The Q-range for this setup is
0.45 to 1.95~\AA$^{-1}$. The data were corrected for detector
efficiency, background container scattering, and self-shielding
using a standard procedure. The Backscattering Neutron measurements
were carried out on IN16 at the Institut Laue-Langevin (Grenoble,
France). The temperature scans have been measured between 2K and
480K at a wavelength of $6.27$~\AA~and the $Q$ ranges from $0.2$ to
$1.9$~\AA$^{-1}$. The data were corrected for detector efficiency,
sample container scattering and self-shielding by the standard
programs. The mean-square displacement is determined from the
observed elastic scattering $I_{el}$ with a resolution
$FWHM=0.85~\mu eV$ (4 ns) which is an average over all scattering
atoms weighted by their cross section. Within the incoherent
Gaussian approximation the effective MSD (relative to the frozen
state) can be deduced from the slope of $\ln(I_{el}/I_{el}(2K))$
versus $Q^2$. The deduced MSD arises for hPS to nearly 92\% from
incoherent scattering of the protons. We measured the partially
deuterated blend sample, where the scattering is dominated (75\%) by
the scattering of the protons, and expect a similar result to that
of a totally hydrogenated blend. The structural characterization of
the blends and pure components have been performed on the D7
spectrometer (ILL) in its diffraction mode, using deuterated samples
to get the coherent information. The incoming wavelength was
$4.8$~\AA~and the investigated Q-range: $0.2$~\AA$^{-1}$ to
$2.75$~\AA$^{-1}$.

\begin{table}[ht!]
\small{
\begin{tabular}{|c|c|c|c|c|}
  \hline
Samples & Mn   & $I_p$=Mw/Mn & $T_g$ (K) & $m_P$\\
 &  ($kg.mol^{-1}$)  &  &  &  \\
\hline
PSh92k& $92$  & $1.04$ & $373.1 \pm 0.4$ & $180 \pm 13$ \\
PSh1790& $1.79$ & $1.06$ & $325.7 \pm 0.2$ & $109 \pm 7$\\
PSh750& $0.735$  & $1.08$  & $276.0 \pm 0.2$  & $113 \pm 5$ \\
PSh92k/PSh750& $1.91$  & $31.3$ &  $325.6 \pm 0.8$  & $85 \pm 6$ \\
PSd110k/PSh750& $1.91$  & $36.1$ & $ 324.7 \pm 0.9$ & $86 \pm 7 $ \\
PSd110k/PSd800& $2.04$ & $33.87$ & $325.3 \pm 0.8$ & $89 \pm 7 $ \\
PSd110k& $109$  & $1.02$ & $ 372.7 \pm 0.4$ & $149 \pm 10 $ \\
PSd2300& $2.21$  & $1.049$ & $336.6 \pm 0.4$ & $129 \pm 3$ \\
PSd800& $0.785$  & $1.10$ & $277.5 \pm 0.2$  & $96 \pm 4$ \\
  \hline
\end{tabular}}
\caption{Sample description and characteristics\label{TableSample}}
\end{table}

\begin{figure}[ht!]
\includegraphics[width=0.9\linewidth]{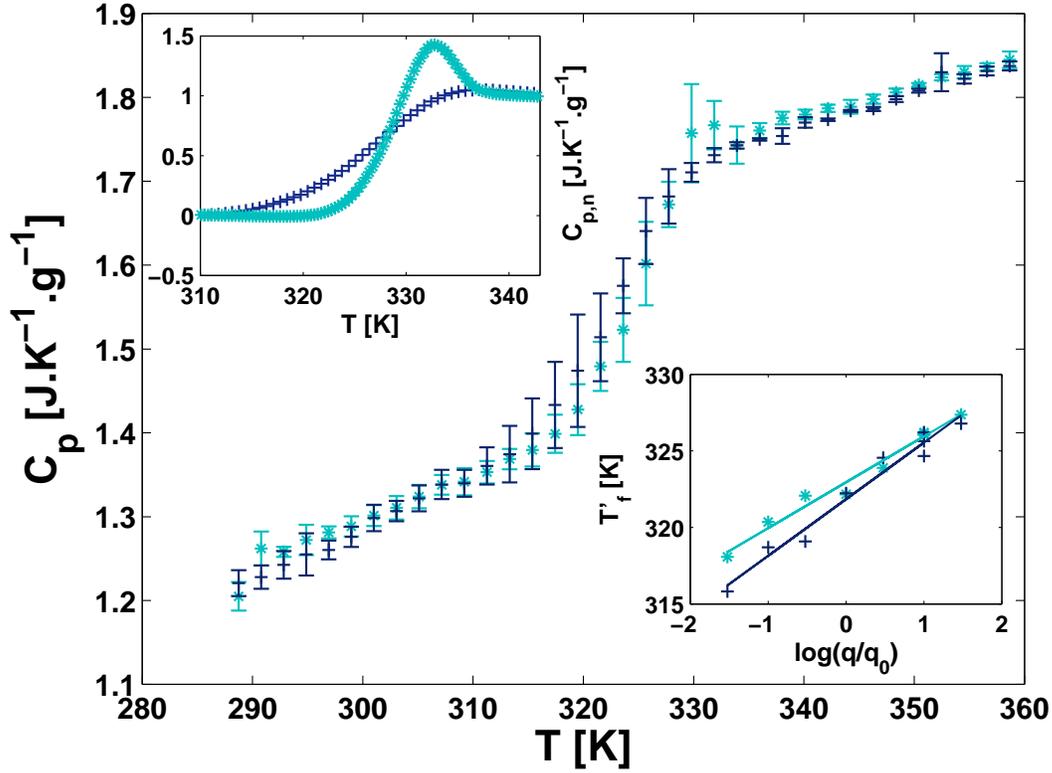}\\
\caption{Absolute heat capacity versus temperature. Monodisperse
PSh1790 (stars) and bidisperse blend PSh92k/h750 (crosses) of same
$T_g$. First inset: Apparent $C_p$ versus T measured by DSC on
heating at 10K.min$^{-1}$ after cooling at 10K.min$^{-1}$. Second
inset: $T$'$_f$ versus logarithm of cooling rate. [View in color for
better clarity].}\label{Cpfig}
\end{figure}

\begin{figure}[ht!]
\includegraphics[width=0.9\linewidth]{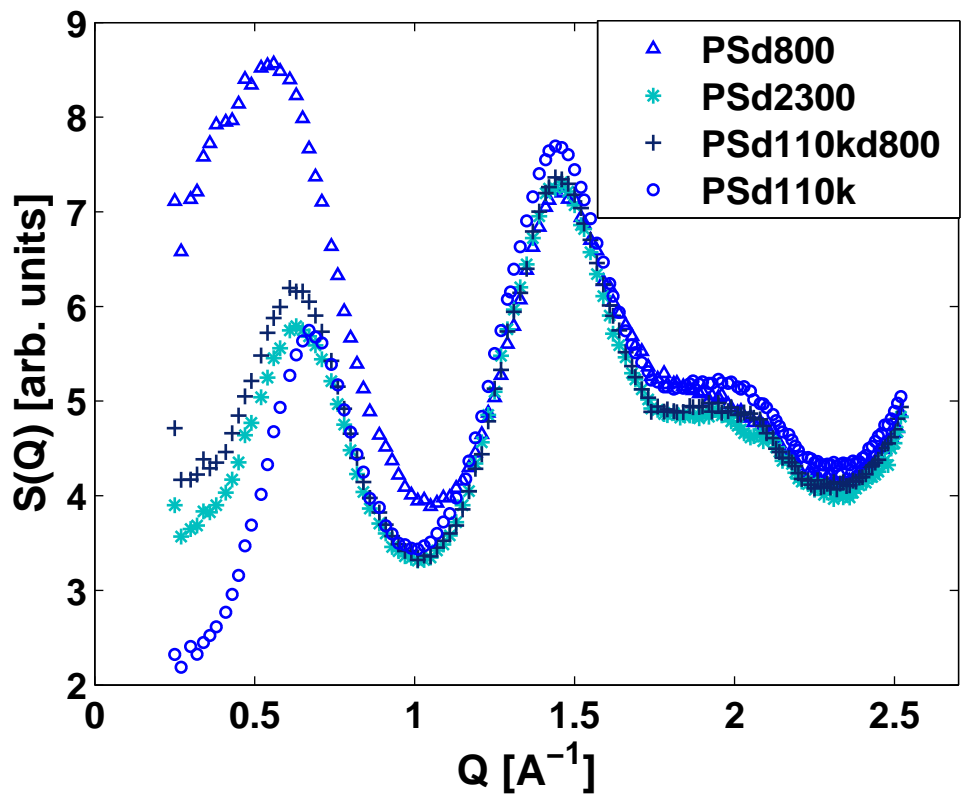}\\
\caption{Structure factor of the samples from the coherent signal of
the deuterated samples at 100K. The pure components of the blend are
also shown to illustrate the $M_n$ dependence of the
structure.}\label{Sqfig}
\end{figure}

\begin{figure}[ht!]
\includegraphics[width=0.9\linewidth]{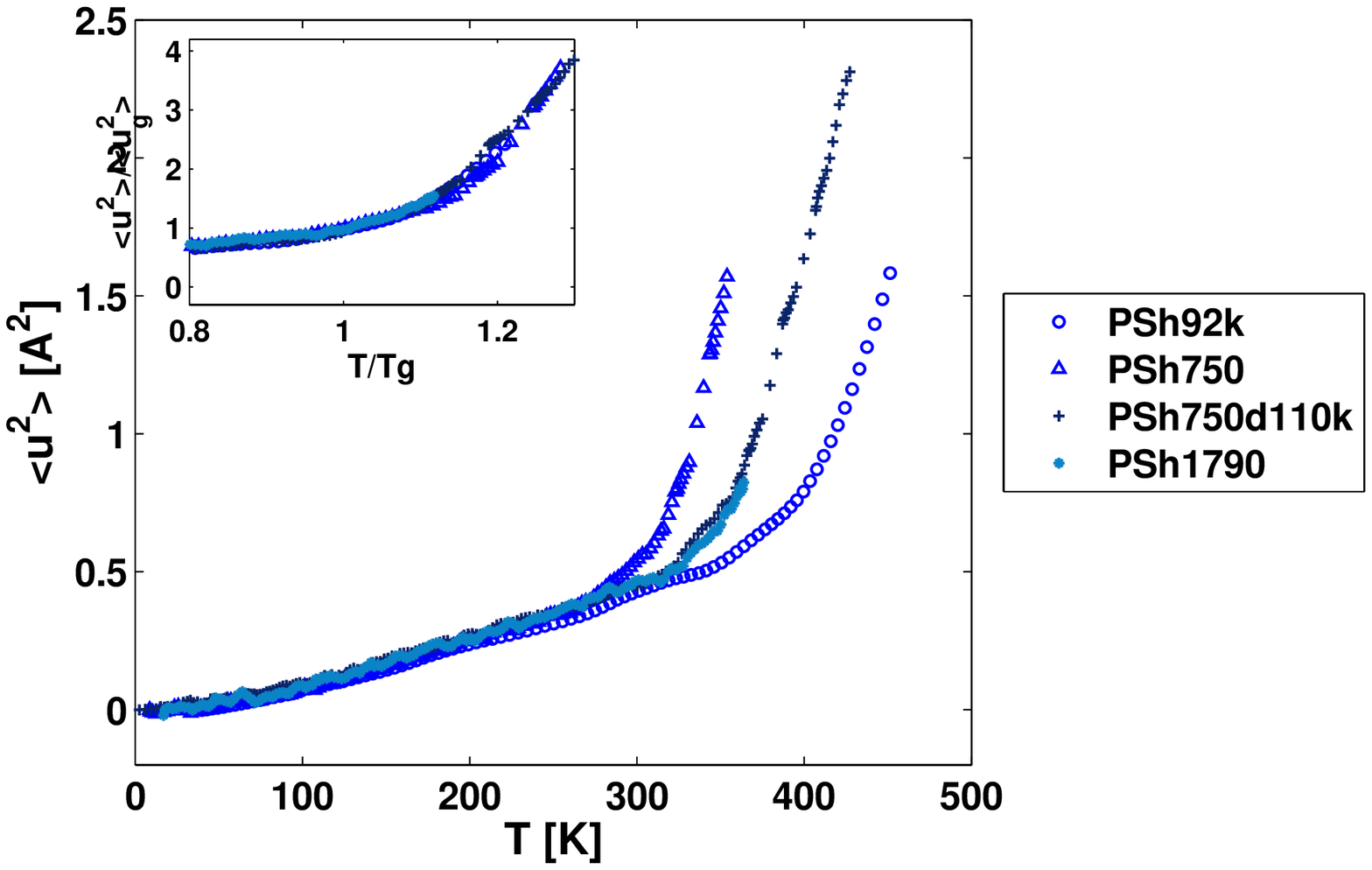}\\
\caption{MSD at the ns timescale of the pure components and the
blend. Inset: rescaled data, $\frac{<u^2>}{<u^2>_g}$ versus
$T/T_g$.}\label{u2fig}
\end{figure}

Figure \ref{Cpfig} shows that the blend and PSh1790 have the same
glass transition temperature and absolute heat capacity as measured
by step scan experiments, both deep in the glass and in the liquid
state. In the first inset, however, the normalized heat capacity
measured on heating shows a different broadening of the glass
transition and aging effects, illustrating that their slow dynamics
differs. The difference in fragility and apparent activation energy
is evidenced by the different dependences of the limiting fictive
temperatures on cooling rate \cite{Wang02}, shown in the second
inset. The fragility obtained for the blend is $20\%$ lower
($m_P$=85) than for the monodisperse PSh1790, $m_P$=109, which is
consistent with previous ones \cite{Privalko86,Robertson04} on other
PS blends. Moreover the blend fragility is closer to that of its low
$M_w$ component and much stronger than the high $M_w$ component. By
fitting the $C_p$ curves with the Tool-Narayanaswamy-Moynihan model
\cite{Moynihan76} of structural recovery, we extracted a Kohlraush
stretching exponent $\beta_{KWW}$ for the blend of $0.4$ lower than
for the 17-mer, of approximatively $0.7$, consistent with the
obvious fact that blending broadens the distribution of relaxation
times. Thus, we have a polymer/oligomer blend with a lower fragility
and a lower $\beta_{KWW}$ than the reference monodisperse sample; on
the other hand, a decrease in fragility is usually accompanied with
an increase in  $\beta_{KWW}$ in pure molecular liquids and polymers
\cite{Bohmer93,Duvvuri04}. In his work on a polydisperse high
molecular weight PS, Privalko \cite{Privalko86} found close results
but he explained it by a change in the local PS structure. Our
results do not support such structural changes as can be seen in
figure \ref{Sqfig}. It shows the structure factor for deuterated
samples at 100K. Two main peaks are observed: the first one near
5~nm$^{-1}$ corresponds to the typical interchain distance
(involving C and D chain-chain correlations), whereas the second
around 13-14~nm$^{-1}$ corresponds to the  distance between phenyl
groups (phenyl-phenyl correlations), as is well known from
literature\cite{Spyriouni07,Iradi04}. The structure factors of the
blend and of the pure 17-mer PS of same $T_g$ are very similar,
especially when compared to the more significant changes produced by
the molecular weight on the pure components of the blend. The local
density of the blend seems to be well predicted by a simple sum rule
from the pure components local densities which confirms the ideal
character of our blend. The interchain lengthscale corresponds to
the static lengthscale at which the signature of the glass
transition is observed, as shown from the temperature dependence of
the first diffraction peak \cite{Frick89,alba98} or from the change
in the temperature-dependence of the MSD. At the same local scale,
the mobility of the atoms can be measured by neutron backscattering
technique via the mean-square displacement averaged over all the
atoms at the ns scale. The upturn in the temperature dependence
coincides with the macroscopic $T_g$, even though the resolution of
the experiments is 4ns. In figure \ref{u2fig}, we show the MSD of
the blend and some pure PS over a wide temperature range. MSD of the
blend and the 17-mer coincide but are distinct from those of the
pure components of the blend due to their differences in $T_g$.
Starting from the correlation between the location of the upturn of
the MSD and $T_g$, the temperature dependence of the MSD around
$T_g$ has been correlated to the fragility in \cite{Ngai00,Dyre04}.
We do not observe the correlation here: the slope is the same for
the blend and the 17-mer, as can be seen from the inset of figure
\ref{u2fig}.

\begin{figure}[ht!]
\includegraphics[width=0.9\linewidth]{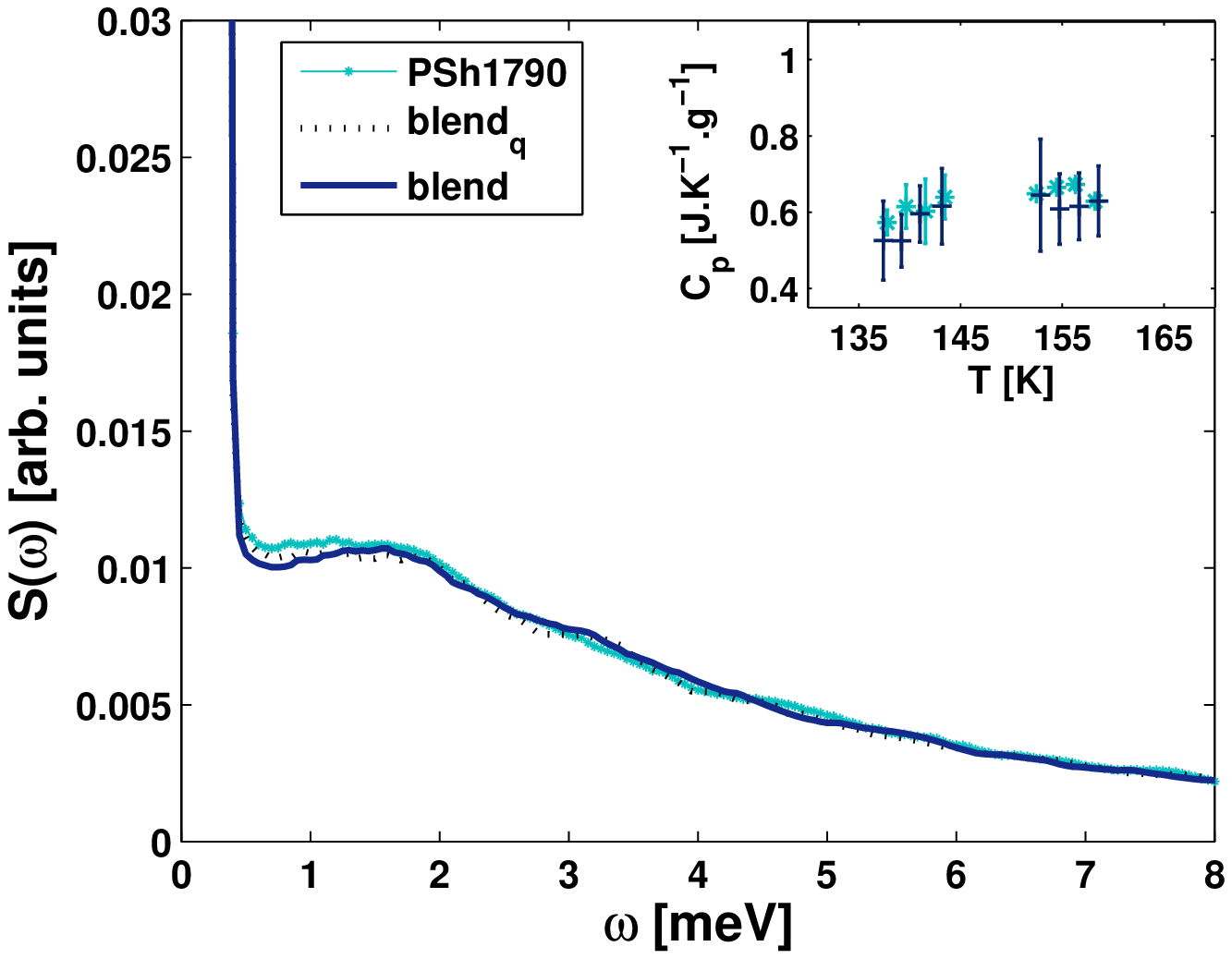}\\
\caption{Dynamic structure factor of the blend and the 17-mer PS at
100K. Inset: low temperature $C_p$ values for both systems measured
by calorimetry.}\label{BPCpfig}
\end{figure}

Another correlation between fast and slow dynamics discussed in the
literature relates the strength of the quasielastic scattering
intensity at $T_g$ (normalized to the intensity of the boson peak as
measured by inelastic scattering) and the fragility
\cite{Sokolov93}. For both the blend and the 17-mer, the dynamic
structure factor is plotted in figure \ref{BPCpfig} at 100K from TOF
neutron scattering experiments: they have superimposable boson peaks
implying similar vibrational density of states (VDOS) consistent
with the very low temperature MSD (figure \ref{u2fig}) and heat
capacity measurements (inset of \ref{BPCpfig}). Only a very small
difference in the quasi-elastic scattering region may be seen, which
is however five times lower than the change observed with molecular
weight (not shown on the figure for reasons of clarity) and could be
attributed here to the thermal treatment we imposed: another
experiment performed on the blend under a rapid quench leads to
almost identical spectra.

One may then wonder what causes the difference in the slow dynamics,
\emph{i.e.} fragility and apparent activation energy of the two
systems. The miscibility of both components in the athermal
polymer/oligomer blend is not enthalpy driven here and must be
accompanied by an increase of entropy. We suggest that a change in
configurational entropy due to mixing is at the origin of the
observed change in the fragility. From recent arguments proposed in
the literature \cite{Pinal08,Schneider97}, the entropy of mixing is
related to the difference between the $T_g$ of the blend calculated
by using simple additivity rules and the measured value. Indeed, we
had to prepare a blend containing $6\%$ less chain ends than in the
monodisperse PSh1790 in order to have the same $T_g$ (it is usually
known in polymeric mixtures that one cannot relate directly the
$T_g$ to the number of chain ends as it is usually done for
monodisperse polymers). The extra mixing contribution to the
configurational entropy can be estimated from recent work
\cite{Pinal08} and is found to be rather small as compared to the
polymer configurational entropy at $T_g$. However, Pinal
\cite{Pinal08} also points out that the cooling rate dependence of
$T_g$ for mixtures implies a cooling rate dependent entropy, which
in turns may be related to the fragility difference between blend
and pure PS.

By taking advantage of polymer molecular weight dependent
properties, we have been able to build a model system that show
direct connection between the slow dynamics and an excess
configurational entropy acting at lengthscales much larger than the
intermolecular (interchain) distance. Two samples of same $T_g$: an
athermal polystyrene/oligomer blend and a reference monodisperse
polystyrene sample were used. They are exactly equivalent at their
local scale, as demonstrated from their structural arrangements and
their fast dynamics and vibrational processes, whereas their slow
dynamics, quantified by the fragilities, are significantly
different. We suggest to rationalize this difference in fragility
via an excess configurational entropy of the polymer/oligomer blend
and establish a direct link between these two quantities. Moreover
we have provided evidence that the origin of the very high fragility
of polymers as compared to molecular liquids involves structural
features occurring at a lengthscale much larger than the interchain
or intermolecular distance. These features are specific to the
polymers and should be taken into account in the development of
theories and models of glass formation.


\end{document}